\documentclass[10pt,letterpaper]{article}

\usepackage[T1]{fontenc}
\usepackage{mathptmx}

\usepackage[letterpaper,left=1.5in,right=1.5in,top=1in,bottom=1in,footskip=0.5in]{geometry}
\usepackage{graphicx}
\usepackage{array}
\usepackage[table]{xcolor}
\usepackage{needspace}
\usepackage{fancyhdr}
\usepackage[hidelinks]{hyperref}
\usepackage{xurl}
\renewcommand{\normalsize}{\fontsize{10}{13}\selectfont}
\normalsize

\definecolor{tblhead}{rgb}{0.878,0.878,0.878}
\definecolor{tblalt}{rgb}{0.961,0.961,0.961}
\definecolor{tblrule}{rgb}{0.667,0.667,0.667}
\newcommand{\bsec}[1]{\par\needspace{4\baselineskip}\vspace{10pt plus 2pt minus 2pt}\noindent{\fontsize{12}{15}\selectfont\textbf{#1}}\par\vspace{-\parskip}\vspace{4pt}\noindent\ignorespaces}
\newcommand{\isub}[1]{\par\needspace{4\baselineskip}\vspace{6pt plus 2pt minus 2pt}\noindent\textbf{#1}\par\vspace{-\parskip}\vspace{3pt}\noindent\ignorespaces}

\begin{document}
\thispagestyle{fancy}
\vspace*{-0.2in}
\noindent\rule{\linewidth}{4pt}\par
\vspace{14pt}
\begin{center}{\fontsize{17}{21}\selectfont\textbf{Trajectories and Comparative Analysis of Global Countries Dominating AI Publications, 2000-2025}}\end{center}
\vspace{2pt}
\noindent\rule{\linewidth}{1pt}\par
\vspace{30pt}
\centerline{\fontsize{11}{13}\selectfont\textbf{Jason Hung}}
\vspace{10pt}
\centerline{\fontsize{10}{12}\selectfont\textit{Pulse Research Fellow, Internet Society}}
\vspace{34pt}
\centerline{\fontsize{12}{15}\selectfont\textbf{Abstract}}
\vspace{8pt}
{\leftskip=0.5in\rightskip=0.5in\noindent This study investigates the shifting global dynamics of Artificial Intelligence (AI) research by analysing the trajectories of countries dominating AI publications between 2000 and 2025. Drawing on the comprehensive OpenAlex datasets and employing fractional counting to avoid double attribution in co-authored work, the research maps the relative shares of AI publications across major global players. The analysis reveals a profound restructuring of the international AI research landscape. The US and the European Union (representing EU27), once the undisputed and established leaders, have experienced a notable decline in relative dominance, with their combined share of publications falling from over 57\% in 2000 to less than 25\% in 2025. In contrast, China has undergone a dramatic ascent, expanding its global share of AI publications from under 5\% in 2000 to nearly 36\% by 2025, therefore emerging as the single most dominant contributor. Alongside China, India has also risen substantially, consolidating a multipolar Asian research ecosystem. These empirical findings raise questions about the strategic implications of concentrated research output, including China\textquoteright{}s potential capacity to shape the future direction of AI innovation, which publication data alone cannot establish. Beyond publication volume, the study further examines citation impact, treated as a proxy for research influence but not research quality, by comparing each country\textquoteright{}s share of high-impact publications against its overall output, and analyses citation impact trajectories across major players. The findings show that in addition to China leading in volume, the country has also recently held the largest share of citation-based high-impact publications. Such an observation challenges the general assumption that Western powers retain dominance in highly cited AI scholarship, although citation-based indicators cannot by themselves establish differences in research quality.\par
\vspace{10pt}\noindent Keywords: artificial intelligence; scientometrics; bibliometric analysis; global research dominance; academic publications; United States; China; European Union\par}
\bsec{1. Introduction}%
The influence of the field of Artificial Intelligence (AI) has rapidly expanded to diverse domains, including global economic competitiveness (Khan et al., 2025), geopolitical relations (Gerlich, 2024), and societal transformation (Gohil, 2023). Its global impact spans sectors from healthcare (Mahdi et al., 2023) and finance (Aldasoro et al., 2024) to defence (Carlo, 2021) and manufacturing (Hong et al., 2025), making the capacity for innovation in AI an impactful indicator of a country\textquoteright{}s future strength and competence. As a result, understanding the global landscape of AI research (specifically, which countries dominate the production of scholarly work and how this dominance evolves over time) is necessary for policymakers, educators, and industry leaders to make informed decisions.

While global leadership in AI is often quantified through metrics such as venture capital investment, the number of successful startups, or the volume of patent applications (Stanford University HAI, n.d.), all known as valuable indicators of commercialisation, the fundamental basis of enduring technological capability remains its academic publication record. Scholarly articles not only record new discoveries and algorithmic breakthroughs but also act as a leading indicator, signalling emerging research directions, highlighting institutional and national strengths, and supplying the open, foundational knowledge upon which future technological development is built. Therefore, a systematic, long-term analysis of the geographical distribution of AI research output, specifically academic publication dominance, provides an important perspective which is often overlooked or underestimated by economic indicators.

This study leverages the comprehensive, open-source OpenAlex datasets to map the shifting dynamics of AI publication dominance across the world\textquoteright{}s leading research countries and regional blocs between 2000 and 2025. In this study, the year 2000 is treated as a meaningful baseline, predating both the rapid deep learning advancement of the early 2010s and China\textquoteright{}s surge in AI publication output. Analysing data between 2000 and 2025 allows the entire transition in global AI publication competition, from Western-led dominance to the present multipolar landscape, to be traced in full. This has created a competitive environment where the historic dominance of established players, such as the US and the European Union (EU), is actively being challenged (Bertelsmann Foundation, 2023). This study offers an opportunity to track the shifting trajectories of these established players alongside the rapid ascent of new powerhouses, most notably China, whose national AI strategy has explicitly prioritised academic output. Through the observation of these simultaneous trajectories, the evolution of a global and fiercely contested research frontier is dissected and comparatively analysed. To be precise, this study aims to satisfy two research aims. First, the longitudinal trajectories of AI publication dominance across major countries and regional blocs between 2000 and 2025 are mapped. Second, whether countries with notable changes in publication volume have proportional variations in citation impact is comparatively analysed, and the predictive (Granger) relationships among these trajectories are examined.

\bsec{2. Literature Review}%
The study of scientific progress and the measurement of national innovation capacity, known as scientometrics, forms the theoretical foundation for this research (Mingers \& Leydesdorff, 2015). In recent decades, bibliometric analysis has become the standard tool for assessing research performance (Hood \& Wilson, 2001). However, applying these methods to highly dynamic, interdisciplinary fields like AI presents unique methodological challenges. The literature review is structured around two key areas, namely the definition and measurement of AI research and the historical and contemporary global landscape of AI production.

\isub{2.1. Defining and Measuring AI Research}%
A major challenge in accurately tracking AI research stems from its multidisciplinary nature. The AI field bridges core computational disciplines, such as computer science and mathematics, and overlaps with engineering, cognitive science, and specialist domains like biomedicine (Abbonato et al., 2024). Consequently, early scientometric studies often relied on narrow, fixed keyword searches or incomplete institutional affiliation lists, a method prone to both exclusion (by missing new subfields) and bias (by over-representing traditional computer science outlets) (Bruce et al., 2025). The consensus in modern scientometric practice, exemplified by approaches adopted by the OECD.ai Observatory, designs the methodology beyond simple keyword approaches. Instead, leveraging comprehensive, curated databases like OpenAlex, which categorise papers based on robust, continuously updated field-of-study taxonomies, is important. This advanced approach, adopted in this research paper, focuses on papers explicitly classified under \textquotedblleft{}AI\textquotedblright{} or \textquotedblleft{}machine learning.\textquotedblright{} While this provides a conservative yet high-confidence measure of core AI research output, this approach ensures consistency and avoids the unreliable categorisation stemming from emerging terminology.

Liu et al. (2021) introduced a bibliometric definition for AI using a hybrid approach, starting by searching core keywords, then extracting high-frequency terms, and compared the outputs against three existing search strategies applied to Web of Science data. Liu et al. (2021) concluded that different search strategies result in substantially different corpus sizes and compositions, while there is no single universally agreed bibliometric definition for AI. Also, F\"{a}rber \& Tampakis (2024) used scientometric data across multiple scholarly databases to compare academic- and company-authored AI publications. Their findings imply that taxonomy choices are contingent on research objectives. Gao et al. (2024) empirically analysed patent-cited AI papers relative to the non-patent-cited counterparts from 1999 to 2013. Their findings show that patent-cited papers have stronger scientific impact, especially for conference publications. Gao et al.\textquoteright{}s (2024) study highlights that not all AI papers are equal in impact.

Furthermore, the literature debates the appropriate counting methodology (Mingers \& Leydesdorff, 2015). Simply counting the total number of published research papers leads to inflation due to increasing co-authorship. Consequently, fractional counting, where credit for a publication is divided equally among the affiliated institutions or countries (as detailed in Section 3.3), has become the gold standard for accurately representing the proportional contribution of each entity (Perianes-Rodr\'{\i}guez et al., 2016). Fractional counting is employed in the methodological design of this research paper.

Publication volume, despite being an indicative metric, fails to fully represent a country\textquoteright{}s scientific advancement in AI. Bibliometric literature distinguishes between quantity and quality, and citation counts are the most widely used proxy for research impact (Hood \& Wilson, 2001), although their validity as a proxy for research quality is limited, since they also indicate field size, citation practices and the visibility of the publishing community. High citation counts do not necessarily follow from high publication volume. A country can dominate in output whilst failing to earn proportional scientific influence, and vice versa (F\"{a}rber \& Tampakis, 2024). Such a quality\textendash{}quantity differentiation is especially obvious for AI publications, as demonstrated by Gao et al. (2024).

\isub{2.2. The Evolving Global Landscape of AI Dominance}%
The narrative of global AI dominance has undergone a seismic shift since the turn of the millennium. Until the mid-2000s, the US and Western Europe, especially the EU bloc, were the leading producers of AI publications (Bertelsmann Foundation, 2023), before China\textquoteright{}s share overtook that of the US between 2005 and 2006 (Figure 4). This early dominance was a direct result of their well-established university systems, robust governmental funding mechanisms dating back to the mid-20th century, and a culture of open academic research (LeCun et al., 2015). The early 2000s were characterised by steady but incremental growth in output among these established players, who benefited from decades of intellectual and infrastructural investment. A comparable trajectory has been documented by Carchiolo and Malgeri (2024), whose bibliometric analysis identifies a crossover between China and the US in AI publication volumes. Such a pattern is verified in Figure 5 of this study. However, Carchiolo and Malgeri\textquoteright{}s (2024) analysis notably relies on Scopus but not OpenAlex, employs raw publication counts but not fractional percentage shares, uses a bespoke set of 18 keywords but not the OECD field-of-study taxonomy, and covers a different timeframe (1995-2023), meaning that the present study is methodologically different from, and complementary to, theirs.

The decline of the West in AI publications can partly be explained structurally. It is noteworthy that following the UK\textquoteright{}s formal departure from the EU in January 2020, this study accounts for the Brexit by treating the UK as an independent trajectory, excluding from the EU27 aggregate at any point, including pre-2020 years. An European Commission report (Balland et al., 2025) finds that European research and innovation hubs are significantly less closely connected than their American counterparts, especially in AI technologies. The report attributes this situation to the structural fragmentation across national borders, such as distinct regulatory regimes, decentralised funding architectures, and limited cross-regional collaboration. In the US, in contrast, Jurowetzki et al. (2021) quantify the growing flow of AI researchers from academia into private technology companies. They discover that researchers specialising in deep learning and those with higher citation impact are most likely to transition from academia to private technology companies. Such a circumstance raises concerns about the privatisation of AI knowledge and the weakening of the public academic research sphere. This circumstance may therefore partly explain why the American academic publication share has declined even though the country remains commercially dominant, since part of this research activity may have migrated out of the academic sphere into industry labs whose outputs are less likely to appear in the OpenAlex corpus. Jurowetzki et al. (2025) second that highly cited researchers from prestigious institutions are relocating to major tech firms, where, upon the transition, their research shows reduced novelty and citation impact.

However, the literature from the past decade highlights the dramatic, centrally-driven emergence of East Asian nations, particularly China (Hamilton-Hart \& Yeung, 2021). Numerous policy analyses and bibliometric reports track China\textquoteright{}s aggressive, sustained investment in AI talent acquisition, large government funding directed by national strategic plans, and infrastructural build-out, all leading to an exponential rise in publication volume (Podda, 2025). This literature suggests that China\textquoteright{}s rise in AI research did not only contribute to the overall global output; rather, China\textquoteright{}s rise challenged the relative dominance of the West (Podda, 2025). The present research, which uses the percentage of global AI publications by country as the key measurement metric, specifically addresses the comparative nature of this global transition, in order to track relative dominance trajectories of given countries. The empirical findings, which demonstrate the US and EU27 shares of global AI publications falling dramatically while China\textquoteright{}s share has surged to over one-third of the global total by 2025, indicate the need for close examination of the underlying competitive dynamics and their geopolitical implications.

\bsec{3. Methodology and Data}%
This study employs a quantitative, data-driven approach to analyse the shifting global dominance in AI publications between the years 2000 and 2025. The analysis focuses on tracking the publication output trajectories of countries to reveal the relative evolution of the global AI research landscape. Sections 3.1 to 3.4 detail the data source, the criteria for identifying AI publications, the counting methodology and the analytical methods employed for both publication quantity and citation impact.

In this study, 12 units of analysis (meaning 11 individual countries and the EU27 as a collective regional bloc) are examined, based on two selection criteria. First, each unit represents a consistently leading contributor to global AI publication output between 2000 and 2025, accounting for the substantial majority of fractional-counted global share in any given year. Second, the selected units correspond to the primary geopolitical actors in contemporary AI governance, strategy and competition discourse, which include the US, China, the EU, India, the UK and other leading Western and Asian economies. The bibliometric analysis is therefore grounded in the broader policy context to which the empirical findings are intended to contribute. Across the 2000 to 2025 period, the selected units jointly account for between approximately 74\% and 86\% of global fractional-counted AI publications in any given year, averaging approximately 80\%. This collective coverage is calculated by summing the shares of the selected units while excluding France, Germany and Spain from the sum, since their output is already included in the EU27 variable, and including the UK, which does not belong to EU27 following its departure from the EU. It should be noted that the UK is treated as a fully independent trajectory throughout this study and is not included within the EU27 aggregate at any point, including pre-2020 years, in order to ensure longitudinal consistency. France, Germany and Spain are reported individually in order to show the trajectories of three major EU member-state contributors, although their output is also included within the EU27 aggregate. Since the HHI (Figure 8), the convergence analysis (Figure 9) and the comparison of high-impact and overall shares (Figure 11a) are computed across all 12 units, the output of these three countries is counted twice in those analyses, once individually and once within the EU27. The series analysed in Figures 10 and 12 include the EU27 but none of its member states individually, although, since these series are renormalised across the 12 units, the overlap still affects them through the denominator used in the renormalisation. The coverage figures reported above and the combined figures in Table 6 count France, Germany and Spain only once, within the EU27. The consequences of this overlap are assessed through a sensitivity check reported in Section 4. For the remaining countries not covered in this study, none individually holds a share large enough to change the comparative trajectories examined here.

\isub{3.1. Data Source and Scope}%
The primary data source for this research is publicly accessible OpenAlex datasets (Priem et al., 2022), downloaded from a comprehensive, open-source bibliographic database. OpenAlex succeeded the Microsoft Academic Graph (Sinha et al., 2015; Wang et al., 2019) and is currently maintained by The OpenResearch Foundation. It provides extensive coverage, encompassing over 245 million research outputs, including journal articles, conference proceedings, and workshop papers. The datasets offer rich bibliographic records, including information about authors, institutions and their corresponding countries, publication venues, and fields of study. Furthermore, all records are tagged with a set of 65,000 topics sourced from Wikidata, and the datasets include citation data, which facilitates the analysis of research impact and citation networks. Its comprehensive nature and interoperability make OpenAlex an ideal foundation for large-scale scientometric studies, although its reference coverage has been shown to diverge from that of the proprietary databases against which OpenAlex is most often benchmarked (Culbert et al., 2025). On OECD.ai, the datasets downloaded are in the form of static csv files. These publicly accessible datasets are not compiled from a live query via the OpenAlex API. Two extractions were used. The descriptive analysis (Figures 1 to 7 and Tables 1 to 7) is based on the OECD.ai series of AI publication shares by country (fractional count) downloaded in late 2025, and the inferential analysis (Figures 8 to 12) is based on the OECD.ai series of AI publication shares, high-impact publications and citations by country downloaded in early 2026. Since OECD.ai notes that OpenAlex data have been affected by snapshot updates and transitions to new data formats (OECD.ai Observatory, n.d.), values for the same year can differ between extractions. In addition, the inferential analyses use shares renormalised across the 12 units, so that the values plotted in Figures 8 to 12 are not directly comparable with the global shares reported in Tables 1 to 7. Institutional affiliation data, on which country attribution depends, are available for approximately 51\% of AI publications in OpenAlex (OECD.ai Observatory, n.d.), and country shares therefore demonstrate only the subset of AI publications for which affiliation information is available. OpenAlex assigns institutional affiliations to countries and territories using ISO 3166-1 codes, under which Taiwan, Hong Kong, and Macau hold codes distinct from that of mainland China. The figures reported here for China therefore represent mainland China only, and are labelled China (mainland) accordingly. It should be noted that publication figures for 2025 may underrepresent the full annual outputs, as indexing delays may occur where work recently published in late 2025 may not be made available within the OpenAlex database. As the descriptive and the inferential analysis outputs were developed in late 2025 and early 2026 respectively, the 2025 data presented in this paper should be interpreted as provisional but not complete.

\isub{3.2. Identification of AI Publications}%
The research scope is focused on the subset of publications relevant to AI, aligning with the criteria utilised by the OECD AI Policy Observatory (OECD.ai). A publication within the OpenAlex datasets is categorised as an AI paper if the publication is tagged during the concept detection operation with a field of study belonging to either the \textquotedblleft{}AI\textquotedblright{} or the \textquotedblleft{}machine learning\textquotedblright{} fields within the OpenAlex taxonomy. It is important to note that results from adjacent fields of study, such as \textquotedblleft{}natural language processing,\textquotedblright{} \textquotedblleft{}speech recognition,\textquotedblright{} and \textquotedblleft{}computer vision\textquotedblright{}, are only included if they also concurrently belong to the aforementioned core \textquotedblleft{}AI\textquotedblright{} or \textquotedblleft{}machine learning\textquotedblright{} classifications (OECD.ai Observatory, n.d.). As this classification relies on rigorous taxonomic assignment, the resultant body of AI publications for the analysis is likely to be conservative, providing a high-confidence set of AI-focused research outputs. Using this taxonomy ensures that the figures reported in this research paper are comparable with the published statistics of the OECD AI Policy Observatory. The taxonomy favours precision over breadth of coverage, although a broader definition would include more applied AI work.

\isub{3.3. Publication Counting Methodology}%
To establish the absolute trajectory of publication output for each country, a fractional counting methodology is employed to avoid the problem of double-counting in co-authored publications. While each publication counts as one unit towards an entity (a country or an institution) in absolute terms, credit for multi-authored papers is distributed equally among the institutions involved. Specifically, a publication written by multiple authors from different institutions is fractionally split among each author based on their institutional affiliation. For instance, if a publication lists four authors affiliated with institutions in the US, one author from an institution in China, and one author from a French institution, the publication is attributed as follows: the US receives 4/6 (four-sixths) of the publication count, China receives 1/6, and France receives 1/6 (OECD.ai Observatory, n.d.). Such counting strategies ensure that the quantity measure accurately captures the distributed contribution of each country. It is important to note that fractional counting does not fully resolve cases where a single author holds simultaneous affiliations across two or more countries. For example, a researcher with joint appointments at institutions in the US and China. In such cases, credit assignment depends on how OpenAlex disambiguates affiliation metadata at the record level, which may not be consistent across all entries in the datasets. Such an approach represents a known limitation of bibliometric analyses relying on institutional affiliation data and is acknowledged as such in the interpretation of empirical findings.

\isub{3.4. Methods}%
All quantitative analyses were conducted via Python (for inferential analysis) and STATA (for descriptive analysis). Multiple datasets were directly downloaded on OECD.ai via OpenAlex and were processed and structured in Python using the pandas library for data manipulation. Descriptive trend analysis and longitudinal trajectory mapping were performed across 2000-2025. Concentration dynamics were then assessed using the Herfindahl\textendash{}Hirschman Index (HHI), with trend significance evaluated through the Mann\textendash{}Kendall non-parametric test and magnitude estimated via Sen\textquoteright{}s slope. Convergence and divergence patterns were further examined through both sigma-convergence ($\sigma$-convergence) and beta-convergence ($\beta$-convergence) regression. Structural breaks in country-level publication trajectories were identified using a single-break Chow test applied to a linear time trend in each country\textquoteright{}s publication share, with the break year selected by maximising the F statistic. Following the OECD.ai classification, a publication is categorised as high-impact if its citation score, calculated as its number of citations divided by the number of years between its year of publication and the upcoming year, exceeds the third quartile of the corresponding distribution of time-normalised citation scores in the AI field (OECD.ai Observatory, n.d.). Although OECD.ai labels publications in this category as high quality, they are referred to in this study as high-impact, because the classification relies solely on citation counts. This study treats the share of high-impact publications and the average citations per paper as proxies for research influence, instead of direct measures of research quality. Citation-based indicators demonstrate the scholarly and industry attention and influence a body of work attracts, which is shaped by field size, citation culture, language, and the cumulative advantage of already-cited work. They do not necessarily indicate methodological rigour, novelty, or reproducibility. The OECD.ai methodology itself notes that citations are an imperfect measure of quality. The high-impact measure applies a time-decay correction, which adjusts for the period since publication, but the measure remains a citation-based proxy. The number of high-impact publications alone is volume-dependent, since a country producing more research papers will tend to produce more high-impact papers in absolute terms. For this reason, Figure 11a compares each country\textquoteright{}s share of high-impact publications against its share of all publications, which measures relative over- or under-performance on impact instead of raw output. Predictive relationships between country trajectories were, in addition, tested using pairwise Granger causality analysis. The figures created for inferential analysis were built using the matplotlib and seaborn libraries in Python, while those for descriptive analysis were made via STATA. Throughout this research paper, statistical significance is reported at the conventional thresholds of p < 0.05, p < 0.01 and p < 0.001, denoted by one, two and three asterisks respectively.

This study specifically uses the datasets displaying the percentage of AI publications instead of the actual number of AI publications by country over time. This means in any given year, the cumulative percentage of AI publications by all countries globally is 100\%. This study decides to use the percentage of AI publications by country metrics as this research paper focuses on comparative data analysis between globally dominant countries. Studying the percentage of AI publications by country longitudinally allows the trajectory of such dominance of a given country to be understood (1) over time and (2) relative to that of other countries.

The HHI was performed as a standard measure of market/system concentration (Rhoades, 1993), calculated by summing the squared publication shares of the 12 units of analysis in a given year after renormalising these shares to sum to 100\%, so that the index ranges from 833, when output is divided equally across the 12 units, to 10,000, when a single unit holds all output. Lower values indicate output dispersed more evenly across the units, and higher values represent concentration in fewer dominant players. Unlike descriptive share trajectories, which show individual country trends, the HHI allows featuring the system-level distribution of AI publication output in a single annual index value. This approach enables an assessment of whether the global landscape as a whole is becoming more or less concentrated over time. The Mann\textendash{}Kendall non-parametric test (Mann, 1945; Kendall, 1975) was then applied to examine whether any trend in the HHI time series is statistically significant, and Sen\textquoteright{}s slope (Sen, 1968) was applied to estimate the magnitude of change per year. Mann\textendash{}Kendall makes no assumption of normality in the time series and is robust to outliers, but the test assumes that successive observations are serially independent, and since annual HHI values are likely to be positively autocorrelated, the reported p-values may overstate statistical significance. The test is applied to the full period as well as to the 2000-2012 and 2013-2025 sub-periods, whose boundary approximately corresponds to the 2012 structural breaks identified in Figure 10, and because this boundary was chosen using the same data, the significant upward trend found for 2013-2025 is interpreted as a description of the period after 2012 but not as confirmation that a new trend began in that year.

The $\sigma$-convergence and $\beta$-convergence were, next, performed as the two complementary approaches drawn from economic growth convergence literature (Barro \& Sala-i-Martin, 1992; Sala-i-Martin, 1996). $\sigma$-convergence measures whether the dispersion of publication shares across the 12 units, measured by the standard deviation of the natural logarithm of their shares in each year, is narrowing or widening over time, with a rising $\sigma$ indicating divergence. $\beta$-convergence tests whether units with lower shares tend to grow faster than those with higher shares, by regressing the annual change in the natural logarithm of each unit\textquoteright{}s share on the natural logarithm of its share in the preceding year, pooled across the 12 units and the 25 annual changes between 2000 and 2025 (N = 300) and estimated by ordinary least squares. A negative and statistically significant $\beta$ would indicate convergence, whereas a $\beta$ indistinguishable from zero indicates the absence of convergence. The pooled specification assumes a common convergence coefficient across units and treats annual changes as conditionally independent, so that the $\beta$ estimate is interpreted as an average tendency across the 12 units but not as the rate at which any individual unit narrows its gap with higher-share units. The convergence framework assesses whether the global AI publication landscape is becoming more equal or polarised across countries. These complementary approaches allow testing whether cross-country variation in publication shares is shrinking or expanding over time, and whether initially lower-share countries are closing the gap with initially higher-share counterparts.

The Chow test (Chow, 1960) was also carried out to statistically identify structural breaks (meaning points in a time series where the underlying trend changes significantly). For each country, a linear time trend in the annual publication share was estimated independently for the periods before and after every candidate break year between 2004 and 2022, which excludes the first and last 15\% of observations as candidate break years, with the candidate year treated as the first year of the post-break regime, and the candidate year yielding the largest Chow F statistic was retained as the estimated break year. A multiple-break procedure such as that of Bai and Perron (1998, 2003) is not applied, because each series contains only 26 annual observations, which leaves too few observations within each regime to estimate and test more than one break with adequate statistical power. Since the break year is selected by maximising the F statistic, conventional F-distribution p-values overstate statistical significance (Andrews, 1993), and significance is therefore also assessed against critical values for the supremum of the F statistic across all candidate break years, obtained by Monte Carlo simulation of 200,000 series of 26 annual observations generated under the null hypothesis of a single linear trend with independent normal errors. Since annual publication shares are serially correlated, which may inflate the test statistics further, each reported break is interpreted as the most pronounced change in a country\textquoteright{}s trend over the study period, and a common break year across multiple countries is treated as indicative, but not conclusive, of a system-level transition in the global AI publication landscape.

The Granger causality analysis was, last but not least, performed to statistically test whether past values of one time series help predict future values of another (Granger, 1969). For each of the 12 directed pairs formed by China, the US, the EU27 and India, a bivariate vector autoregression of the two units\textquoteright{} annual publication shares, renormalised across the 12 units, was estimated, and an F-test of the joint significance of the lagged values of the predicting unit in the equation for the predicted unit was performed. The descriptive and aforementioned inferential analyses establish what has changed in the global AI publication landscape, and the Granger causality analysis evaluates whether knowing a country\textquoteright{}s trajectory in prior years improves prediction of another country\textquoteright{}s trajectory, beyond what the latter\textquoteright{}s historical record already suggests. Granger causality establishes temporal precedence and predictive association but not causation, and the test cannot by itself explain why such changes have occurred. Three further features of the design constrain interpretation. First, each series contains only 26 annual observations (i.e., 2000-25), which limits statistical power. Second, 12 directed pairs are tested without correction for multiple comparisons, so that at the 5\% significance level approximately 0.6 false positives (12 tests $\times$ 0.05 = 0.6) would be expected by chance alone. Third, publication shares are compositional, since a rise in one unit\textquoteright{}s share necessarily reduces the combined share of the others, which can induce predictive relationships that arise from this accounting constraint instead of from any behavioural interdependence.

\bsec{4. Empirical Findings}%
In this research, the focus is placed on major global players in AI publications in Europe, Asia and North America. Figure 1 and Table 1 show the AI publications over year for selected European countries. EU27 refers to the 27 member states of the EU. It is noteworthy that the UK left the EU on 31st January 2020 after Brexit. The UK is treated as a fully independent trajectory throughout this study and is not included within the EU27 aggregate at any point, including pre-2020 years, to ensure longitudinal consistency. The data show that the EU27 accounted for 29.53\% of global AI publications in 2000. The share dropped to some 22\% to 23\% between 2005 and 2015. Over the last decade, such figures have continued to drop significantly, to 12.40\% in 2025. Among Western European countries, the UK and Germany have been leading AI publications from 2000 to 2025.

\par\vspace{6pt}\noindent\begin{minipage}{\linewidth}
\noindent\textbf{Figure 1: Publications Over Year for Selected European Countries}\par
\vspace{4pt}\noindent\includegraphics[width=\linewidth]{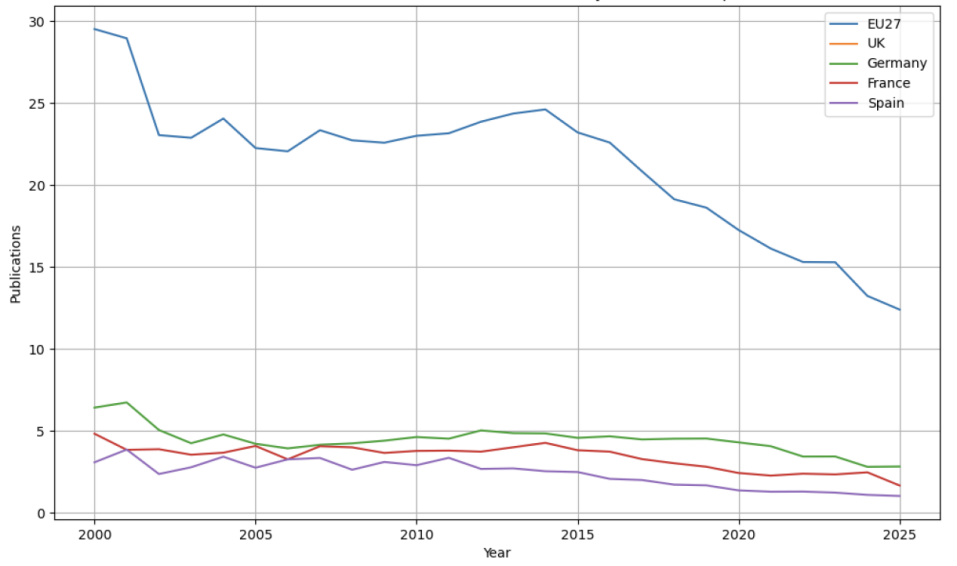}
\end{minipage}\par\vspace{6pt}
\par\vspace{6pt}\noindent\begin{minipage}{\linewidth}
\noindent\textbf{Table 1: Publications Over Year for Selected European Countries}\par\vspace{4pt}
{\setlength{\tabcolsep}{0pt}\renewcommand{\arraystretch}{1.25}\arrayrulecolor{tblrule}\fontsize{9}{11}\selectfont
\noindent\begin{tabular}{|>{\centering\arraybackslash}p{1.450in}|>{\centering\arraybackslash}p{0.667in}|>{\centering\arraybackslash}p{0.667in}|>{\centering\arraybackslash}p{0.667in}|>{\centering\arraybackslash}p{0.667in}|>{\centering\arraybackslash}p{0.667in}|>{\centering\arraybackslash}p{0.667in}|}\hline
\rowcolor{tblhead}\textbf{Country\_label} & \textbf{2000} & \textbf{2005} & \textbf{2010} & \textbf{2015} & \textbf{2020} & \textbf{2025} \\ \hline
EU27 & 29.53 & 22.27 & 23.01 & 23.22 & 17.26 & 12.40 \\ \hline
\rowcolor{tblalt}France & 4.82 & 4.08 & 3.78 & 3.82 & 2.43 & 1.66 \\ \hline
Germany & 6.42 & 4.21 & 4.62 & 4.57 & 4.30 & 2.82 \\ \hline
\rowcolor{tblalt}Spain & 3.08 & 2.75 & 2.90 & 2.49 & 1.37 & 1.02 \\ \hline
United Kingdom & 8.03 & 5.10 & 3.93 & 4.32 & 4.09 & 2.74 \\ \hline
\end{tabular}}
\end{minipage}\par\vspace{6pt}
Figure 2 and Table 2 show the AI publications over year for selected Asian countries. In Asia, Japan used to be a very major global player in AI publications in the 2000s. Its contribution to global AI publications, however, has declined since 2010s. In contrast, China only contributed to just below 5\% of global AI publications in 2000. In 2005, the figure soared to 18.29\%. After declining from 28.66\% in 2010 to 17.02\% in 2015, China\textquoteright{}s share of global AI publications rose again, reaching about 36\% in 2025. Over the last decade, other than China, India has become the leading Asian player in contributing to global AI publications (about 10\% of all AI publications are authored by Indian researchers in 2025).

\par\vspace{6pt}\noindent\begin{minipage}{\linewidth}
\noindent\textbf{Figure 2: Publications Over the Year for Selected Asian Countries}\par
\vspace{4pt}\noindent\includegraphics[width=\linewidth]{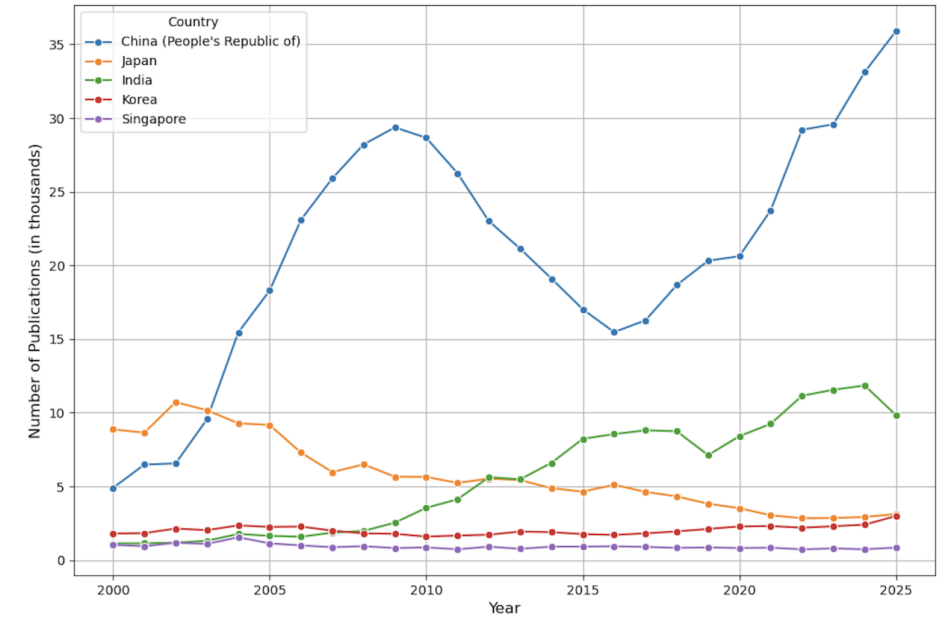}
\end{minipage}\par\vspace{6pt}
\par\vspace{6pt}\noindent\begin{minipage}{\linewidth}
\noindent\textbf{Table 2: Publications Over the Year for Selected Asian Countries}\par\vspace{4pt}
{\setlength{\tabcolsep}{0pt}\renewcommand{\arraystretch}{1.25}\arrayrulecolor{tblrule}\fontsize{9}{11}\selectfont
\noindent\begin{tabular}{|>{\centering\arraybackslash}p{1.450in}|>{\centering\arraybackslash}p{0.667in}|>{\centering\arraybackslash}p{0.667in}|>{\centering\arraybackslash}p{0.667in}|>{\centering\arraybackslash}p{0.667in}|>{\centering\arraybackslash}p{0.667in}|>{\centering\arraybackslash}p{0.667in}|}\hline
\rowcolor{tblhead}\textbf{Country\_label} & \textbf{2000} & \textbf{2005} & \textbf{2010} & \textbf{2015} & \textbf{2020} & \textbf{2025} \\ \hline
China & 4.90 & 18.29 & 28.66 & 17.02 & 20.63 & 35.91 \\ \hline
\rowcolor{tblalt}India & 1.13 & 1.64 & 3.55 & 8.23 & 8.41 & 9.84 \\ \hline
Japan & 8.87 & 9.17 & 5.65 & 4.65 & 3.53 & 3.13 \\ \hline
\rowcolor{tblalt}Korea & 1.80 & 2.25 & 1.60 & 1.76 & 2.29 & 2.99 \\ \hline
Singapore & 1.04 & 1.14 & 0.86 & 0.92 & 0.82 & 0.85 \\ \hline
\end{tabular}}
\end{minipage}\par\vspace{6pt}
Figure 3 and Table 3 show the AI publications over year for selected North American countries. In North America, Canada\textquoteright{}s contribution to AI publications has dropped by half between 2000 (3.32\%) and 2025 (1.67\%). The US, which remained the larger North American contributor throughout the study period, has also seen its percentage of AI publications cut by half between 2000 and 2025, from 27.56\% to 12.01\%.

\par\vspace{6pt}\noindent\begin{minipage}{\linewidth}
\noindent\textbf{Figure 3: Publications Over Year for Selected North American Countries}\par
\vspace{4pt}\noindent\includegraphics[width=\linewidth]{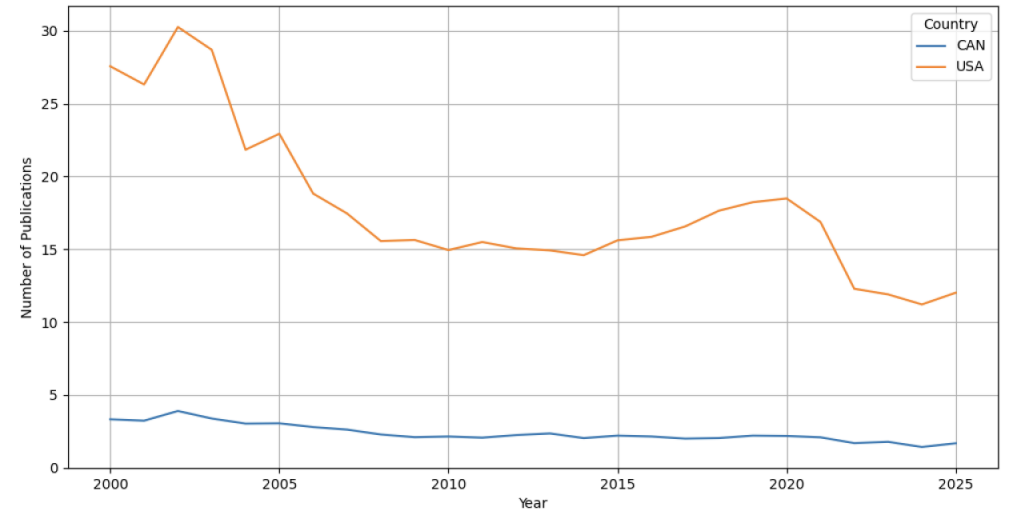}
\end{minipage}\par\vspace{6pt}
\par\vspace{6pt}\noindent\begin{minipage}{\linewidth}
\noindent\textbf{Table 3: Publications Over Year for Selected North American Countries}\par\vspace{4pt}
{\setlength{\tabcolsep}{0pt}\renewcommand{\arraystretch}{1.25}\arrayrulecolor{tblrule}\fontsize{9}{11}\selectfont
\noindent\begin{tabular}{|>{\centering\arraybackslash}p{1.450in}|>{\centering\arraybackslash}p{0.667in}|>{\centering\arraybackslash}p{0.667in}|>{\centering\arraybackslash}p{0.667in}|>{\centering\arraybackslash}p{0.667in}|>{\centering\arraybackslash}p{0.667in}|>{\centering\arraybackslash}p{0.667in}|}\hline
\rowcolor{tblhead}\textbf{Country\_label} & \textbf{2000} & \textbf{2005} & \textbf{2010} & \textbf{2015} & \textbf{2020} & \textbf{2025} \\ \hline
Canada & 3.32 & 3.05 & 2.14 & 2.20 & 2.18 & 1.67 \\ \hline
\rowcolor{tblalt}United States & 27.56 & 22.92 & 14.94 & 15.61 & 18.48 & 12.01 \\ \hline
\end{tabular}}
\end{minipage}\par\vspace{6pt}
Figure 4 and Table 4 show the AI publications over year between the US and China. These empirical outputs allow the trajectories of percentage of AI publications by the world\textquoteright{}s two leading forces to be comparatively analysed. The data show that in 2000, the US dominated 27.56\% of global AI publications, which was over five times higher than the 4.90\% from China. Between 2005 and 2006, China\textquoteright{}s global AI publication contributions reached the level of the US. From 2006 to 2016, China\textquoteright{}s contributions to global AI publications outnumbered those of the US. Their contributions to global AI publications maintained at very close levels between 2016 and 2017. However, since 2020, China\textquoteright{}s dominance in AI publications has surged, while the US\textquoteright{}s dominance has declined. In 2025, China has contributed three times as many AI publications as the US (35.91\% vs 12.01\%).

\par\vspace{6pt}\noindent\begin{minipage}{\linewidth}
\noindent\textbf{Figure 4: Publications Over Time for United States and China}\par
\vspace{4pt}\noindent\includegraphics[width=\linewidth]{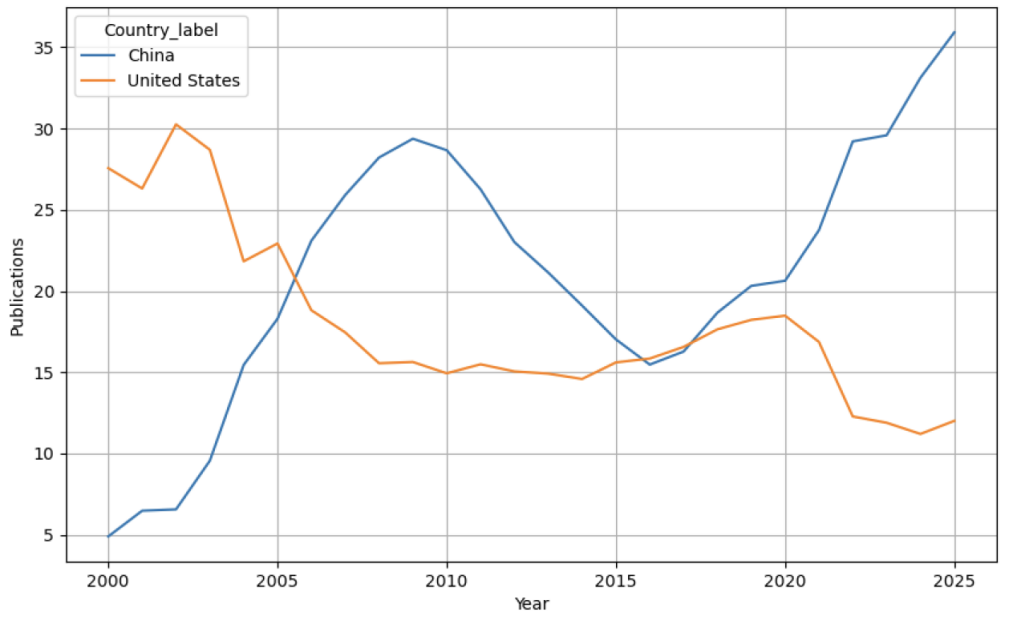}
\end{minipage}\par\vspace{6pt}
\par\vspace{6pt}\noindent\begin{minipage}{\linewidth}
\noindent\textbf{Table 4: Publications Over Time for United States and China}\par\vspace{4pt}
{\setlength{\tabcolsep}{0pt}\renewcommand{\arraystretch}{1.25}\arrayrulecolor{tblrule}\fontsize{9}{11}\selectfont
\noindent\begin{tabular}{|>{\centering\arraybackslash}p{1.450in}|>{\centering\arraybackslash}p{0.667in}|>{\centering\arraybackslash}p{0.667in}|>{\centering\arraybackslash}p{0.667in}|>{\centering\arraybackslash}p{0.667in}|>{\centering\arraybackslash}p{0.667in}|>{\centering\arraybackslash}p{0.667in}|}\hline
\rowcolor{tblhead}\textbf{Country\_label} & \textbf{2000} & \textbf{2005} & \textbf{2010} & \textbf{2015} & \textbf{2020} & \textbf{2025} \\ \hline
China & 4.90 & 18.29 & 28.66 & 17.02 & 20.63 & 35.91 \\ \hline
\rowcolor{tblalt}United States & 27.56 & 22.92 & 14.94 & 15.61 & 18.48 & 12.01 \\ \hline
\end{tabular}}
\end{minipage}\par\vspace{6pt}
Figure 5 and Table 5 supplement the preceding empirical outputs by showing the AI publications over year between the US, China and EU27. The data show that in 2000, the US and EU27 were the two leading powerhouses in contributing to global AI publications. In 2005, China joined the US and EU27 to become one of the clear-cut leading global players in AI publications. The shares of all three players remained comparatively close between 2015 and 2020. Since 2020, the contributions to AI publications by the US and EU27 has declined consistently, while China\textquoteright{}s dominance has reached new heights.

\par\vspace{6pt}\noindent\begin{minipage}{\linewidth}
\noindent\textbf{Figure 5: Publications Over Time for United States, China, and EU27}\par
\vspace{4pt}\noindent\includegraphics[width=\linewidth]{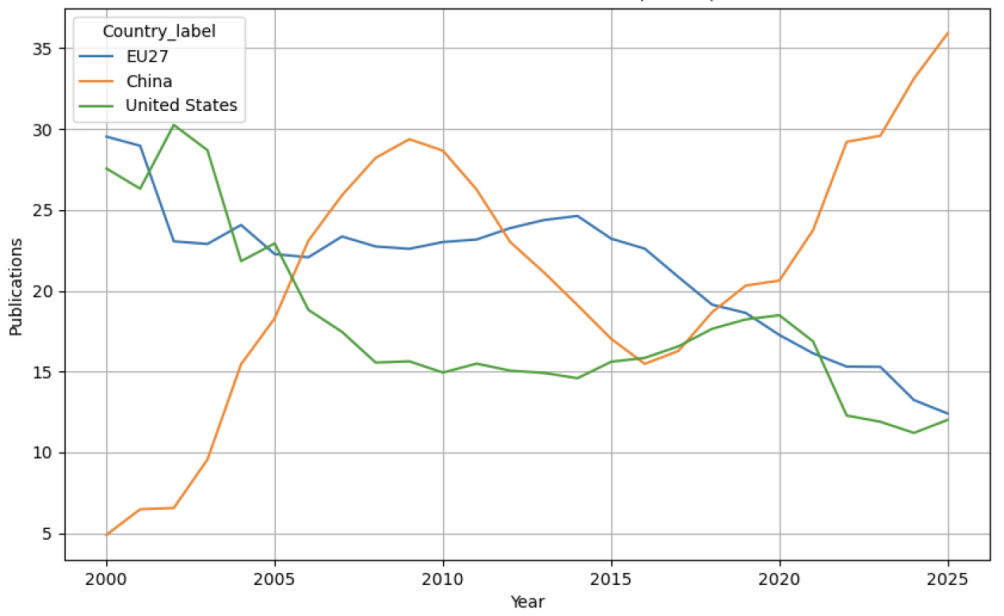}
\end{minipage}\par\vspace{6pt}
\par\vspace{6pt}\noindent\begin{minipage}{\linewidth}
\noindent\textbf{Table 5: Publications Over Time for United States, China, and EU27}\par\vspace{4pt}
{\setlength{\tabcolsep}{0pt}\renewcommand{\arraystretch}{1.25}\arrayrulecolor{tblrule}\fontsize{9}{11}\selectfont
\noindent\begin{tabular}{|>{\centering\arraybackslash}p{1.450in}|>{\centering\arraybackslash}p{0.667in}|>{\centering\arraybackslash}p{0.667in}|>{\centering\arraybackslash}p{0.667in}|>{\centering\arraybackslash}p{0.667in}|>{\centering\arraybackslash}p{0.667in}|>{\centering\arraybackslash}p{0.667in}|}\hline
\rowcolor{tblhead}\textbf{Country\_label} & \textbf{2000} & \textbf{2005} & \textbf{2010} & \textbf{2015} & \textbf{2020} & \textbf{2025} \\ \hline
China & 4.90 & 18.29 & 28.66 & 17.02 & 20.63 & 35.91 \\ \hline
\rowcolor{tblalt}EU27 & 29.53 & 22.27 & 23.01 & 23.22 & 17.26 & 12.40 \\ \hline
United States & 27.56 & 22.92 & 14.94 & 15.61 & 18.48 & 12.01 \\ \hline
\end{tabular}}
\end{minipage}\par\vspace{6pt}
Figure 6 and Table 6 show the AI publications over time between China and non-China global players combined. Non-China global players combined refers to all EU27 countries, the US, Canada, the UK (where applicable), Japan, Korea, Singapore and India, with France, Germany and Spain counted only once, within the EU27. The data show that all these major global players (including China) have combined for over 78\% of AI publications annually between 2021 and 2025. In 2021, China\textquoteright{}s contributions were fewer than half of those from all other major global players combined (23.74\% vs 54.32\%). Yet, China\textquoteright{}s share has continued to grow, while the combined share of all non-China major global players fell from 54.32\% in 2021 to 45.64\% in 2025, with a slight increase only between 2022 and 2023. In 2025, China\textquoteright{}s contributions to global AI publications reached 35.91\%, while all non-China global players combined for 45.64\%.

\par\vspace{6pt}\noindent\begin{minipage}{\linewidth}
\noindent\textbf{Figure 6: Publications Over Time for China and Non-China Global Players Combined}\par
\vspace{4pt}\noindent\includegraphics[width=\linewidth]{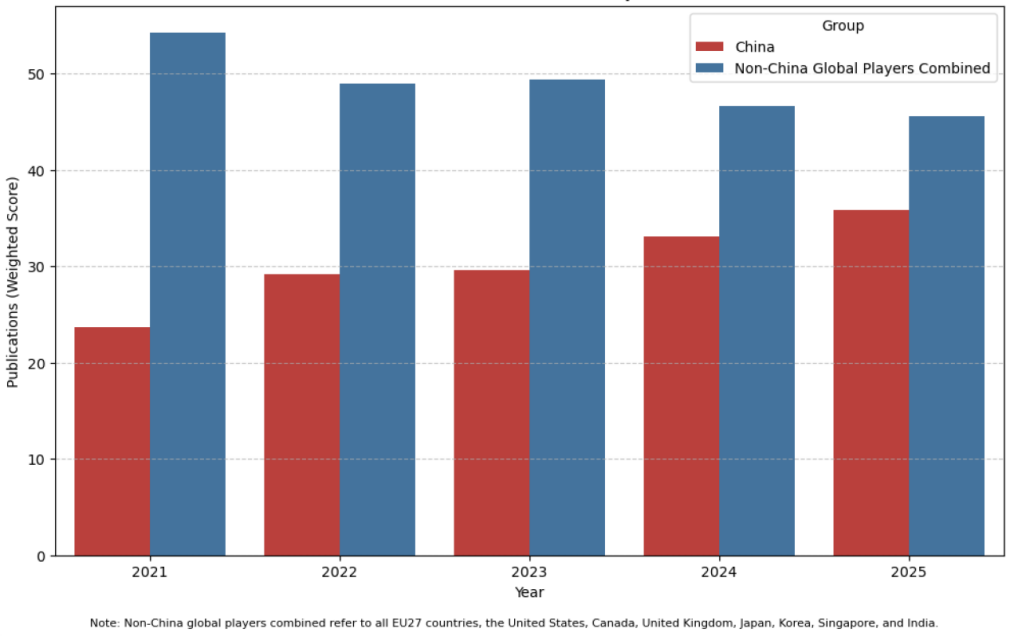}
\end{minipage}\par\vspace{6pt}
\par\vspace{6pt}\noindent\begin{minipage}{\linewidth}
\noindent\textbf{Table 6: Publications Over Time for China and Non-China Global Players Combined}\par\vspace{4pt}
{\setlength{\tabcolsep}{0pt}\renewcommand{\arraystretch}{1.25}\arrayrulecolor{tblrule}\fontsize{9}{11}\selectfont
\noindent\begin{tabular}{|>{\centering\arraybackslash}p{2.200in}|>{\centering\arraybackslash}p{0.650in}|>{\centering\arraybackslash}p{0.650in}|>{\centering\arraybackslash}p{0.650in}|>{\centering\arraybackslash}p{0.650in}|>{\centering\arraybackslash}p{0.650in}|}\hline
\rowcolor{tblhead}\textbf{Country\_label} & \textbf{2021} & \textbf{2022} & \textbf{2023} & \textbf{2024} & \textbf{2025} \\ \hline
China & 23.74 & 29.20 & 29.58 & 33.12 & 35.91 \\ \hline
\rowcolor{tblalt}Non-China Global Players Combined & 54.32 & 49.02 & 49.45 & 46.61 & 45.64 \\ \hline
\end{tabular}}
\end{minipage}\par\vspace{6pt}
Figure 7 and Table 7 show the year-over-year (YOY) percentage change in publications over year between the US, China and EU27. The data show that China\textquoteright{}s YOY percentage change resulted in a boom during the mid-2000s. China experienced another YOY percentage change boom in the early 2020s. China\textquoteright{}s YOY increases in 2004, 2006 and 2022 coincided with the three largest YOY decreases in the US, although China\textquoteright{}s increases and US decreases do not coincide in every year, as in 2018, when both China and the US recorded positive YOY changes. The coincidence of these changes aligns with the fact that publication shares sum to 100\% in every year, since a gain in one unit\textquoteright{}s share must be offset by a combined loss across all other units.

\par\vspace{6pt}\noindent\begin{minipage}{\linewidth}
\noindent\textbf{Figure 7: Year-Over-Year Percentage Change in Publications for United States, China, EU27}\par
\vspace{4pt}\noindent\includegraphics[width=\linewidth]{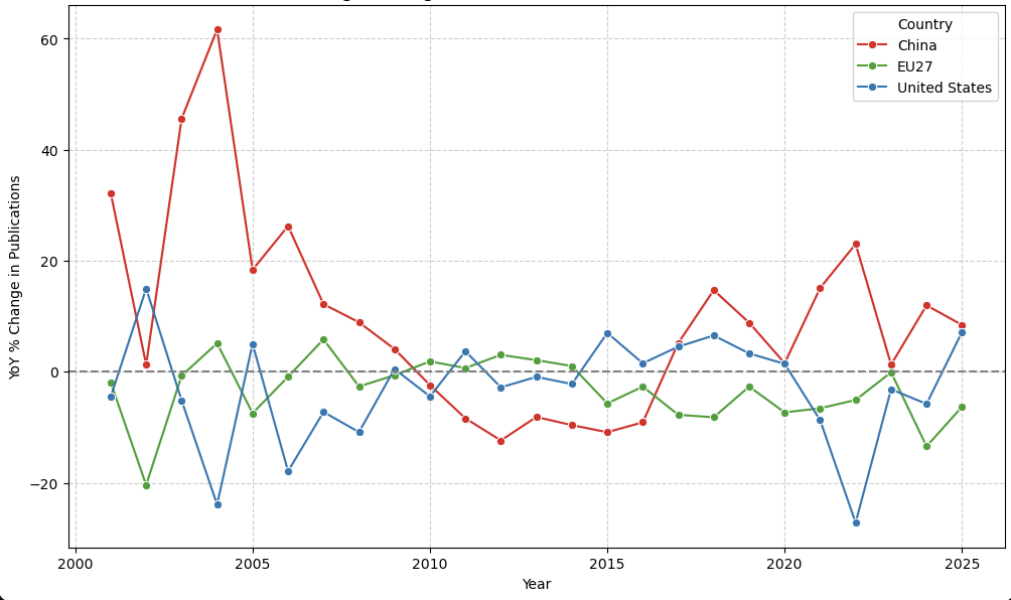}
\end{minipage}\par\vspace{6pt}
\par\vspace{6pt}\noindent\begin{minipage}{\linewidth}
\noindent\textbf{Table 7: Year-Over-Year Percentage Change in Publications for United States, China, EU27}\par\vspace{4pt}
{\setlength{\tabcolsep}{0pt}\renewcommand{\arraystretch}{1.25}\arrayrulecolor{tblrule}\fontsize{9}{11}\selectfont
\noindent\begin{tabular}{|>{\centering\arraybackslash}p{1.450in}|>{\centering\arraybackslash}p{0.800in}|>{\centering\arraybackslash}p{0.800in}|>{\centering\arraybackslash}p{0.800in}|>{\centering\arraybackslash}p{0.800in}|>{\centering\arraybackslash}p{0.800in}|}\hline
\rowcolor{tblhead}\textbf{Country\_label} & \textbf{2005} & \textbf{2010} & \textbf{2015} & \textbf{2020} & \textbf{2025} \\ \hline
China & 18.38 & -2.40 & -10.89 & 1.54 & 8.43 \\ \hline
\rowcolor{tblalt}EU27 & -7.48 & 1.87 & -5.70 & -7.32 & -6.29 \\ \hline
United States & 5.02 & -4.44 & 6.99 & 1.42 & 7.18 \\ \hline
\end{tabular}}
\end{minipage}\par\vspace{6pt}
Figure 8 presents the HHI analysis of global AI publication concentration across the study period. The index does not follow a monotonic path, since the HHI fluctuated in the early 2000s, rose to a local peak in 2009, fell to its lowest value in 2016, and then grew to its highest values in 2024 and 2025. Accordingly, the Mann\textendash{}Kendall test finds no statistically significant trend over the full period ($\tau$=0.169, p=0.234) or in the 2000-2012 sub-period ($\tau$=0.231, p=0.300), whereas the 2013-2025 sub-period shows a statistically significant upward trend ($\tau$=0.667, p=0.002), with Sen\textquoteright{}s slope estimating concentration increasing at approximately +63 HHI units per year, compared with +8.16 HHI units per year over the full period. The finding complements the individual country trajectories presented in Figures 1-7, by highlighting the system-level distributional transition that single-country graphs fail to display. AI publication output among the 12 units has become more concentrated since 2013, the time period in which the deep learning breakthrough (LeCun et al., 2015) and the intensification of China\textquoteright{}s national AI investment strategy took place, and this increase in concentration is driven predominantly by the growth of China\textquoteright{}s share, not by the consolidation of several dominant players.

\par\vspace{6pt}\noindent\begin{minipage}{\linewidth}
\noindent\textbf{Figure 8: Market Concentration of AI Publication Shares (HHI), 2000-2025, among 12 Major Global AI-Publishing Countries}\par
\vspace{4pt}\noindent\includegraphics[width=\linewidth]{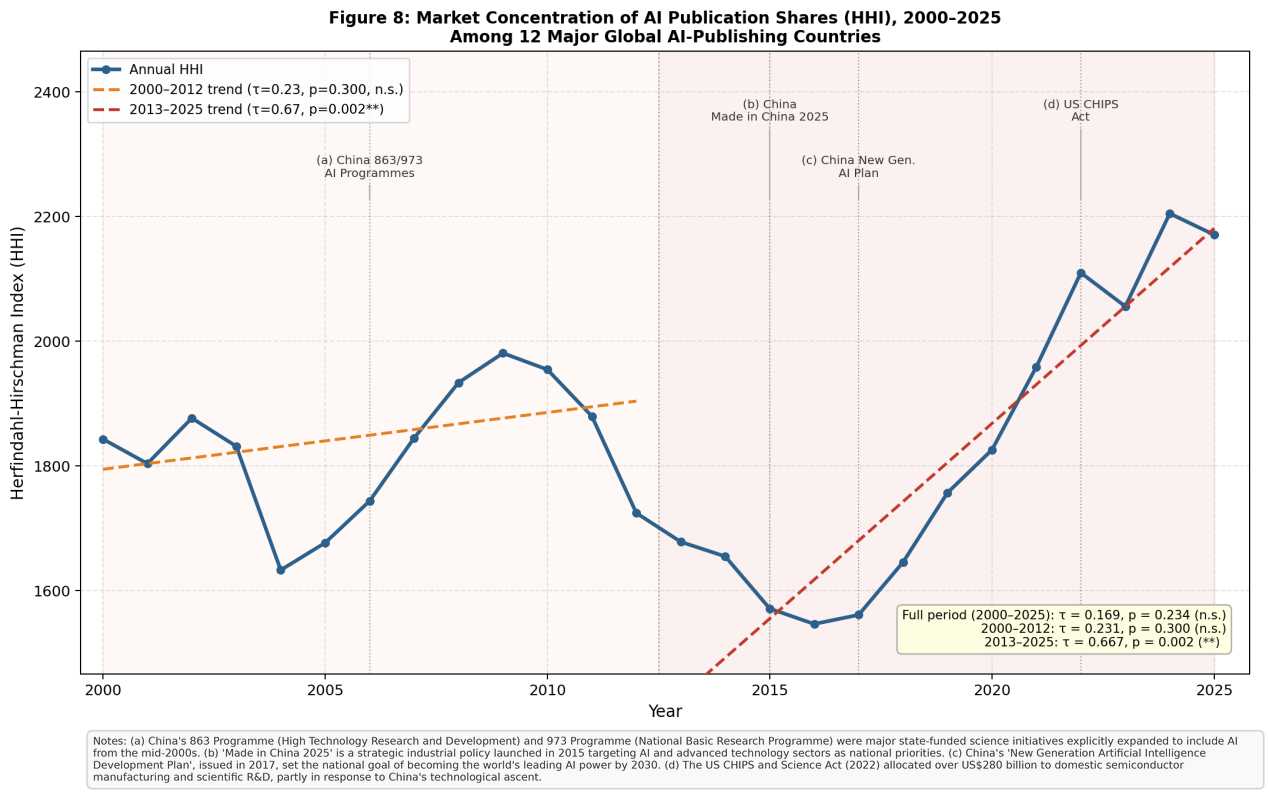}
\end{minipage}\par\vspace{6pt}
Figure 9 presents the convergence analysis of AI publication shares across the study period. The $\sigma$-convergence results show that the standard deviation of log publication shares remained broadly stable between 2000 ($\sigma$=0.987) and 2012 ($\sigma$=0.979), fell to its lowest value in 2015 (approximately $\sigma$=0.94), and then rose to $\sigma$=1.155 in 2025, indicating that dispersion across the 12 units has widened since the mid-2010s. The $\beta$-convergence regression yields a coefficient that is close to zero and statistically insignificant ($\beta$=0.0007, SE=0.0064, p=0.912, R\textsuperscript{2}<0.001, N=300), meaning that a unit\textquoteright{}s log share in one year carries no information about the change in its log share in the following year. This finding indicates that countries with initially lower publication shares have not increased their shares at a faster rate than those with initially higher shares. Overall, the empirical findings shown in Figure 9 suggest that dispersion among the 12 units has widened since the mid-2010s, with no indication of $\beta$-convergence over the full period. The absence of $\beta$-convergence should not, however, be taken to mean that dominant units are entrenched, since the two largest units in 2000, the US and the EU27, both lost more than half of their shares over the study period, so that the widening of dispersion results from the rise of China relative to the other units, not from the consolidation of incumbent leaders. As a sensitivity check on the overlap between the EU27 and its member states, the HHI and $\sigma$ (computed as the population standard deviation of log shares) were recalculated from the five-year values reported in Tables 1 to 3, first across all 12 units and then across nine units after excluding France, Germany and Spain, and since these values come from the descriptive extraction, they differ in level from the annual values plotted in Figures 8 and 9. Under both specifications, the HHI reaches its lowest five-year value in the second half of the 2010s (1,561 in 2015 with 12 units and 1,948 in 2020 with nine units) and its highest value in 2025 (2,278 and 2,578 respectively), and $\sigma$ rises after 2015 (from 0.940 to 1.092 with 12 units and from 1.055 to 1.112 with nine units), although with nine units $\sigma$ in 2025 remains below its 2000 value of 1.179. The rise in concentration and dispersion since the mid-2010s is therefore robust to the treatment of the overlap, whereas the comparison of dispersion between 2000 and 2025 is not, and the divergence finding is accordingly confined to the period since the mid-2010s.

\par\vspace{6pt}\noindent\begin{minipage}{\linewidth}
\noindent\textbf{Figure 9(a) (Left): $\sigma$-Convergence (Cross-Country) Dispersion of AI Publication Shares}\par
\noindent\textbf{Figure 9(b) (Right): $\beta$-Convergence (Initial Share vs. Subsequent Growth Rate, 2000-2025)}\par
\vspace{4pt}\noindent\includegraphics[width=\linewidth]{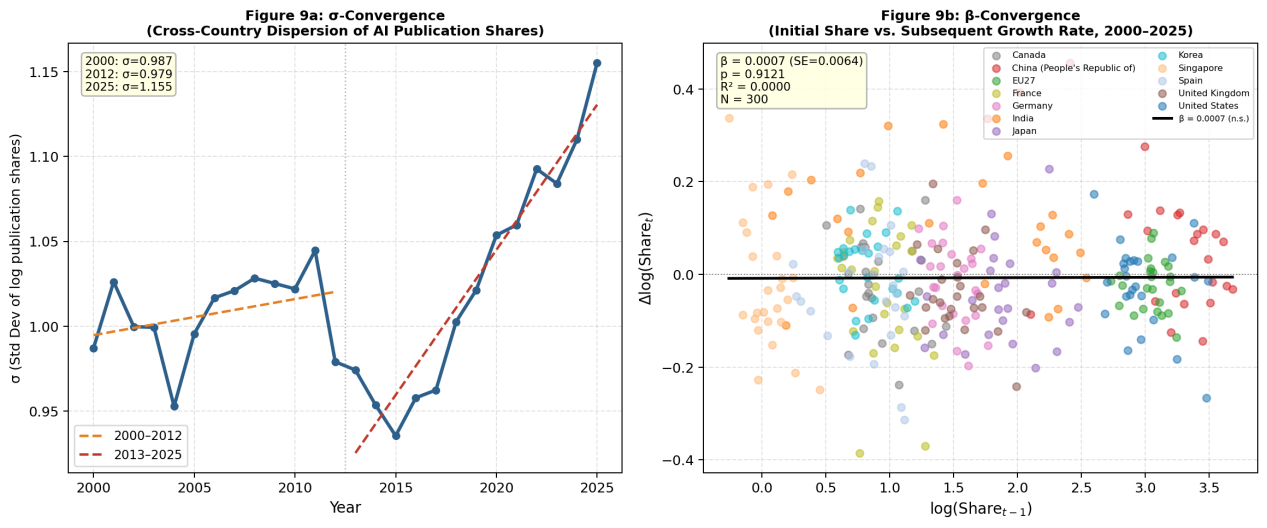}
\end{minipage}\par\vspace{6pt}
Figure 10 presents the Chow test (Chow, 1960) structural break detection results for five countries across the 2000-2025 period. Three countries share a common structural break year of 2012, namely China (F=29.5), EU27 (F=28.8) and India (F=11.4), while the US (F=25.7) and Japan (F=8.8) share a different common break year of 2006. Against the simulated critical values for the supremum F statistic, the breaks for China, the EU27 and the US are statistically significant at the 0.1\% level, the break for India at the 1\% level, and the break for Japan at the 5\% level, whereas the conventional F-distribution p-values (p<0.001 for China, the EU27, India and the US, and p<0.01 for Japan) overstate the significance of the breaks for India and Japan. The coincidence of break years across China, EU27 and India at 2012 suggests a common transition point, which coincides in timing with the deep learning breakthrough and the intensification of China\textquoteright{}s national AI investment strategy, although the Chow test establishes only when a trajectory changes and not why the trajectory changes. For China, the 2012 break divides the rapid expansion of its share up to its first peak in 2009 from a subsequent phase in which its share declined until 2016 and then rose to the highest values of the series, so that the break signals the end of China\textquoteright{}s first expansion phase instead of the beginning of its current leading position. The earlier 2006 break for the US and Japan identifies the point at which both countries\textquoteright{} shares shifted to a lower level, and neither country has since recovered its pre-2006 share.

\par\vspace{6pt}\noindent\begin{minipage}{\linewidth}
\noindent\textbf{Figure 10: Structural Break Detection in AI Publication Share Trajectories (Chow Test, 2000-2025; Significance Markers Based on Conventional F-Distribution p-Values)}\par
\vspace{4pt}\noindent\includegraphics[width=\linewidth]{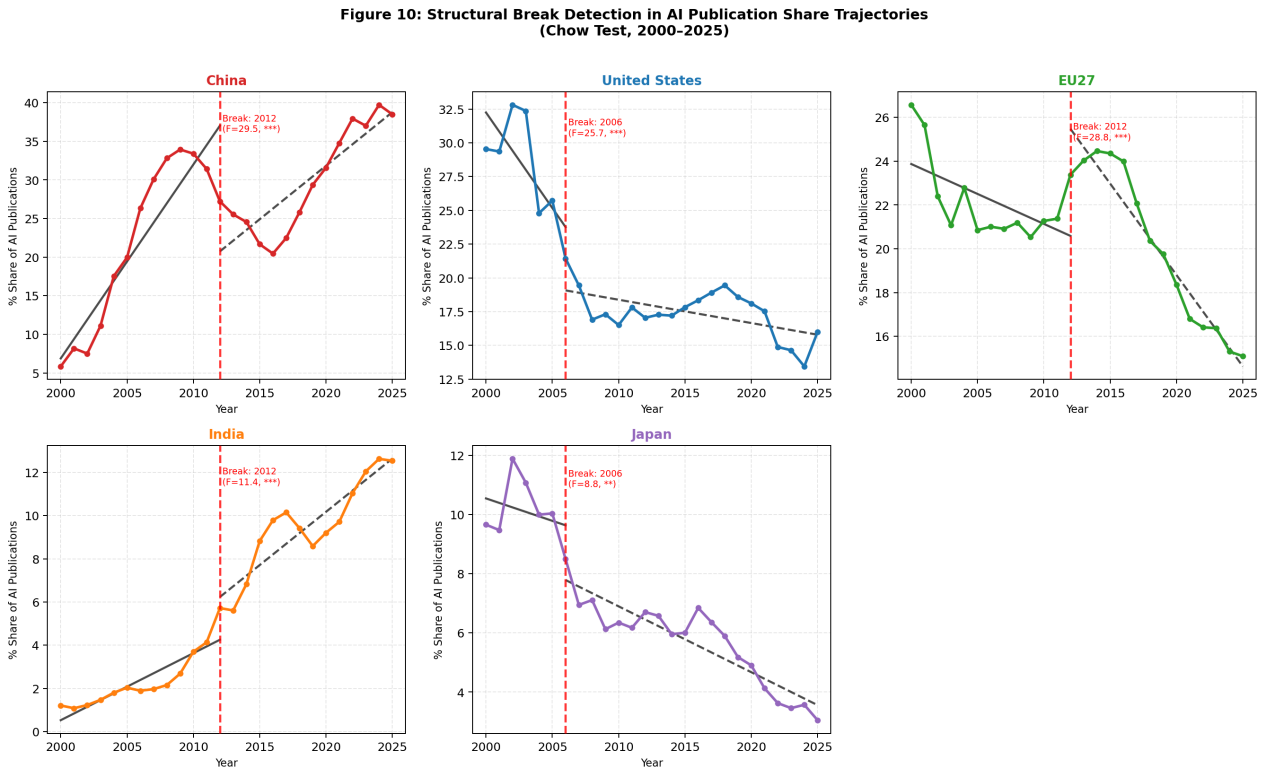}
\end{minipage}\par\vspace{6pt}
Figure 11 presents the comparison of publication volume and citation impact in 2025 along with citation impact trajectories across the study period. The axes in Figure 11a show each unit\textquoteright{}s share within the 12 units analysed in this study (11 individual countries and the EU27 bloc), renormalised to sum to 100\% across these units, but not its share of total global output reported in Tables 1 to 7, and since France, Germany and Spain are included both individually and within the EU27, their output is counted twice in this renormalisation. Figure 11a shows that China is the only unit positioned clearly above the diagonal on which the high-impact share equals the all-publications share, with its share of high-impact publications (approximately 48\%) substantially exceeding its share of all publications (approximately 38\%). These findings indicate that China\textquoteright{}s output is disproportionately concentrated in highly cited work relative to its volume. The US and EU27 both sit slightly below the diagonal, with shares of high-impact publications of approximately 13\% to 14\% against shares of all publications of approximately 15\% to 16\%, while India sits clearly below the diagonal (approximately 8\% against approximately 13\%). Figure 11b reports the average number of citations per AI paper for each unit and year, as provided in the OECD.ai citation data derived from OpenAlex. The US recorded the highest value among the units shown until around 2018, except in 2016, when the UK briefly exceeded the US value, and the US value peaked at approximately 69 citations per paper in 2001 before falling to approximately 15 by 2025. China\textquoteright{}s value, in contrast, rose to approximately 38 citations per paper by 2025, apart from a temporary dip in 2016, and has exceeded that of the US since around 2019. China therefore holds the largest share of high-impact AI publications as defined by the OECD.ai citation-based classification, and its share of high-impact output exceeds its share of overall output, which indicates that China over-performs on citation impact relative to its publication volume. Citation counts measure the attention that a body of work receives, which is shaped by field size, the size and citation practices of national research communities, language, and the cumulative advantage of already-cited work, and they do not directly measure methodological rigour, novelty, or reproducibility. Figure 11 therefore challenges the general assumption that Western powers retain dominance in highly cited AI scholarship, but the figure alone does not establish that Chinese AI research is of higher quality, since citation-based measures proxy influence over quality.

\par\vspace{6pt}\noindent\begin{minipage}{\linewidth}
\noindent\textbf{Figure 11: Publication Volume and Citation Impact of AI Research}\par
\vspace{4pt}\noindent\includegraphics[width=\linewidth]{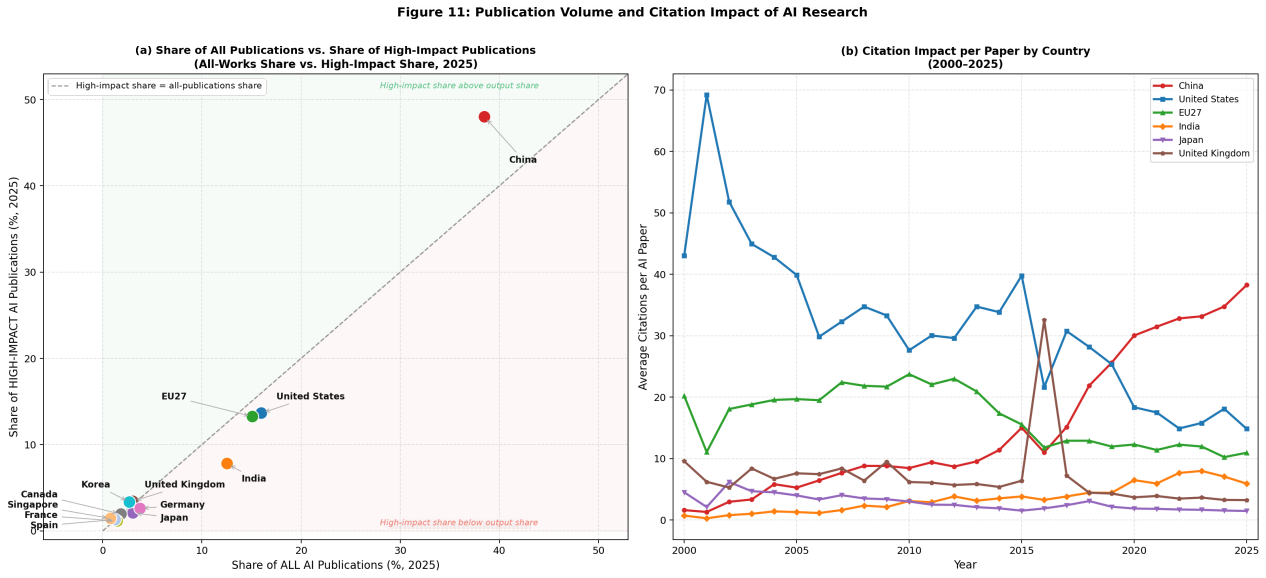}
\end{minipage}\par\vspace{6pt}
Figure 12 presents the Granger causality analysis of AI publication share dynamics across China, the US, EU27 and India. Out of the 12 directed relationships tested in Figure 12(a), only two are statistically significant at the 5\% level, namely from the US to China (p = 0.017) and from the EU27 to the US (p = 0.039), while the relationship from China to the US is marginal (p = 0.061). Past values of the US publication share improve the prediction of China\textquoteright{}s subsequent trajectory beyond what China\textquoteright{}s own historical values predict alone. Likewise, past values of the EU27 publication share improve the prediction of the US\textquoteright{}s subsequent trajectory beyond what the US\textquoteright{}s own historical values predict alone. No statistically significant predictive relationships are found among the remaining pairs, and none of the relationships involving India is statistically significant. Neither statistically significant relationship would remain significant after a Bonferroni correction for 12 tests, which lowers the threshold to p < 0.0042. The empirical findings therefore offer only limited support for the view that the past trajectory of the US share carries predictive information for China\textquoteright{}s subsequent trajectory, and they do not support a general relationship running from Western to Asian trajectories, since the second statistically significant relationship links two Western units and no relationship involving India is statistically significant. It is noteworthy that Granger causality tests for statistical association and temporal precedence but not causation.

\par\vspace{6pt}\noindent\begin{minipage}{\linewidth}
\noindent\textbf{Figure 12: Granger Causality Analysis of AI Publication Share Dynamics}\par
\vspace{4pt}\noindent\includegraphics[width=\linewidth]{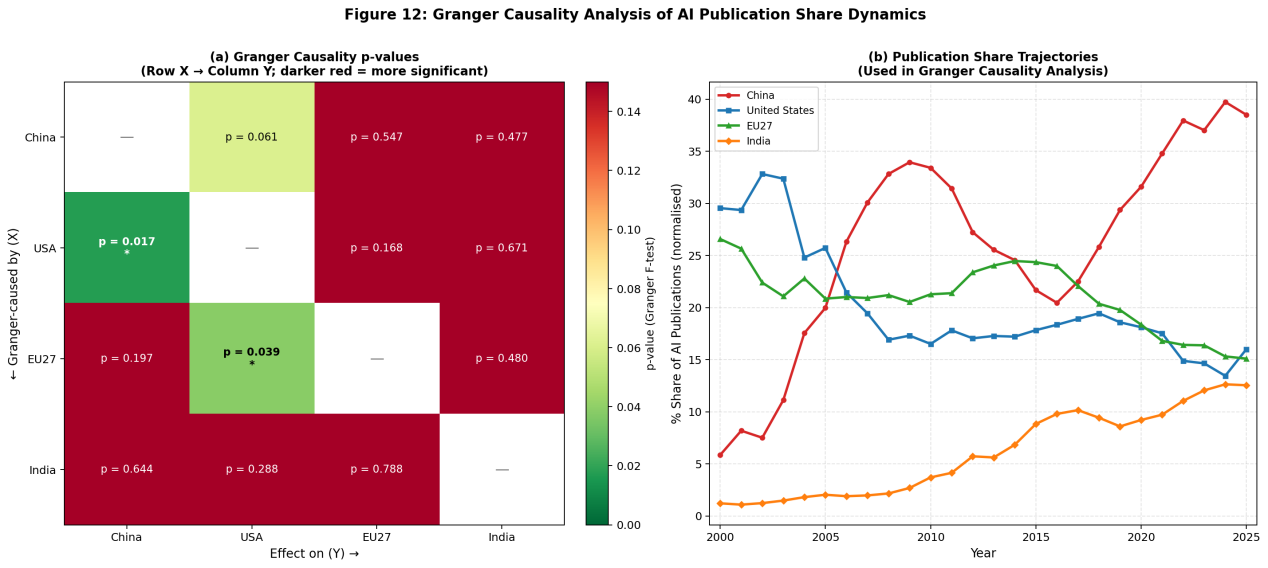}
\end{minipage}\par\vspace{6pt}
\bsec{5. Discussion}%
The empirical findings of this study reveal a profound restructuring of the global AI research landscape between 2000 and 2025. By using the metric of percentage of global AI publications by country, this research helps map the relative dominance trajectories of major global players. The data demonstrate a substantial change in the global distribution of AI publications, characterised by the long-term decline in the relative shares of the historic Western powerhouses (the US and the EU27) and the rise of China, which was interrupted by a decline in China\textquoteright{}s share in the first half of the 2010s.

\isub{5.1. The Decline of Western Dominance}%
The empirical findings for the US and the EU27 illustrate a clear decline of their collective global dominance in AI publications. In 2000, these two entities commanded a combined share of global AI publications exceeding 57\% (27.56\% for the US and 29.53\% for the EU27). By 2025, their combined share has plummeted to approximately 24.41\%. There are two thematic observations that can be made.

First, the US, while initially maintaining a high global share of AI publications, has seen its relative share reduced by more than half, falling to 12.01\% in 2025. This decline, particularly noticeable during China\textquoteright{}s mid-2000s and post-2020 surge periods (Figure 7 \& Table 7), suggests that while US AI research remains influential, recording the second-highest average citations per paper among the units shown in Figure 11b in 2025, its capacity to grow its share of the volume of foundational AI research is diminishing relative to global growth, particularly in Asia, led by China and India.

Second, the EU27 exhibits an even steeper and more consistent decline, dropping from nearly 30\% to 12.40\% over the period (Figure 1 \& Table 1). Key individual member states have also declined over the same period, with Germany falling from 6.42\% to 2.82\% and France from 4.82\% to 1.66\%, while the UK, which is excluded from the EU27 throughout this study, fell from 8.03\% to 2.74\%. This decline aligns with arguments that the fragmentation of European research and innovation systems limits their capacity to compete with the centralised national strategies of other regions, such as East Asia (Balland et al., 2025; Hamilton-Hart \& Yeung, 2021), although this study does not examine whether such fragmentation causes the decline.

\isub{5.2. The Ascendancy of China and Asia, and Its Implications}%
The metric used in this study indicates relative dominance. This does not mean the actual volumes of AI publications from the US and the EU27 have been declining. Instead, the empirical findings show the fall in global dominance from Western powers is associated with the ascendancy of Asian powers, led by China, in AI publications. The most notable finding is the dramatic change in China\textquoteright{}s trajectory in AI publications, which the literature attributes to national AI policy (Choi and Yoon, 2025). Sitting at a modest 4.90\% in global share in 2000, China surpassed the US between 2005 and 2006 and the EU27 by 2010, fell back below the EU27 in 2015 (17.02\% against 23.22\%), and has led every other unit since the late 2010s, reaching nearly 36\% of the global total by 2025. First, China\textquoteright{}s rapid growth, with large positive YOY percentage changes in the mid-2000s and early 2020s (Figure 7 \& Table 7), aligns with accounts attributing its expansion to centralised strategic investment in academic output (Podda, 2025), although this study does not examine whether such investment causes the expansion. This leading position in publication volume indicates that the largest single share of the world\textquoteright{}s AI publications now originates from Chinese institutions (Ellis, 2025).

Second, the empirical findings indicate a narrowing gap between China and the other major global players combined, since China\textquoteright{}s share rose from 23.74\% in 2021 to 35.91\% in 2025, while the combined share of the US, EU27, the UK, Canada, Japan, Korea, Singapore and India fell from 54.32\% to 45.64\% over the same period (Figure 6 \& Table 6). This narrowing gap may carry geopolitical implications, since the literature argues that control over the production of foundational AI knowledge can translate into strategic advantages in setting technical standards, developing proprietary technologies, and securing future economic competitiveness (Quimba \& Barral, 2024). This study does not examine these implications, since the analysis measures shares of academic publications but not technological or strategic capability.

Third, while today\textquoteright{}s global focus remains on the US-China dynamics (Figure 4 \& Table 4), the significant growth of India, whose share has grown from 1.13\% to nearly 10\% by 2025, signals the rise of a diversified, multipolar Asian research ecosystem. The combined share of AI publications by non-China Asian powers (India, Japan, Korea and Singapore) (Figure 2 \& Table 2) exceeded that of the EU27 and the UK combined (Figure 1 \& Table 1) in 2025 (16.81\% against 15.14\%), having remained below that combined share in 2020 (15.05\% against 21.35\%). This indicates that the decline in Western shares is attributable not only to China but also to the wider growth of publication share across Asia.

\isub{5.3. Structural Concentration and Polarisation of Global AI Research}%
In addition to the geographical redistribution documented in Figure 1-7, the HHI (Figure 8) and convergence (Figure 9) analyses demonstrate a structural dimension of the global AI publication landscape that is not displayed when studying volume trajectories alone. The HHI analysis shows that AI publication output among the 12 units has become more concentrated since 2013, with a statistically significant upward trend in the 2013-2025 sub-period according to the Mann\textendash{}Kendall test ($\tau$=0.667, p=0.002), although no statistically significant trend is found over the full 2000-2025 period. The convergence analysis reinforces this finding, since the standard deviation of log publication shares across the 12 units has widened since the mid-2010s, and the $\beta$-convergence regression finds no statistically significant tendency for lower-share countries to close the gap with dominant players. These findings indicate that AI publication output among the 12 units has become more polarised since the mid-2010s, although this polarisation is driven predominantly by the rise of a single unit, China, but not by the accumulation of advantages among incumbent leaders, since the US and the EU27 both lost share over the same period. These findings therefore align only partly with the cumulative advantage dynamics documented in the scientometric literature (Merton, 1968; Mingers \& Leydesdorff, 2015). Such findings carry policy implications. If the gap between dominant and non-dominant players continues to widen at the current rate, the likelihood of a more balanced distribution of global AI research output diminishes over time. For Western policymakers, the challenge is therefore not only to reverse the relative decline in publication share, but also to address the structural conditions that the literature associates with this decline, such as research funding fragmentation in the EU (Balland et al., 2025) and the migration of high-impact researchers from academia to industry in the US (Jurowetzki et al., 2021, 2025), although these conditions are not examined directly in this study.

\isub{5.4. China\textquoteright{}s Position from Publication Volume to Citation Impact}%
The comparison of publication volume and citation impact presented in Figure 11 qualifies the narrative of Western decline in AI scientific output. The general assumption in academic and policy discourse has been that Western countries, especially the US, dominate highly cited AI scholarship, despite China leading in publication volume. Figure 11(a) challenges this general assumption, since China is the only unit whose share of high-impact publications substantially exceeds its share of all publications in 2025, while the US and the EU27 sit slightly below parity. Figure 11(b) further indicates that China\textquoteright{}s average citations per paper has grown over the study period and has exceeded that of the US since around 2019, whereas the US value fell from a peak of approximately 69 citations per paper in 2001 to approximately 15 by 2025. These findings indicate that China\textquoteright{}s leading position extends from publication volume to citation-based high-impact output. They do not, however, establish that Chinese AI research is of higher quality, since citation-based indicators also depend on the size and citation practices of national research communities. The migration of high-impact researchers from Western academic institutions to private industry documented by Jurowetzki et al. (2021) and Jurowetzki et al. (2025), along with the fragmentation of EU research investment argued by Balland et al. (2025), offers a plausible, though untested, explanation of why Western citation impact has declined while China\textquoteright{}s has increased.

The Granger causality analysis presented in Figure 12 offers a further, but limited, perspective on global AI publication dynamics. The finding that past values of the US publication share improve the prediction of China\textquoteright{}s subsequent trajectory aligns with an interpretation in which shifts in US publication activity preceded adjustments in China\textquoteright{}s research system. Such responsiveness would align with accounts of East Asian states adapting national strategy to transitions in global markets and competition (Hamilton-Hart \& Yeung, 2021), and with the policy-driven expansion of Chinese AI research output documented by Choi and Yoon (2025). Another interpretation is that research priorities and methodological frameworks developed in Western academic contexts, most notably the deep learning methods that reshaped the field from the early 2010s (LeCun et al., 2015), were subsequently adopted and scaled by Asian institutions. These interpretations remain speculative, because the relationship is no longer statistically significant after correction for multiple comparisons, because publication shares sum to 100\% in every year, so that a change in one unit\textquoteright{}s share necessarily changes the shares of the others, and because the Granger causality test establishes temporal precedence and predictive association but not a causal relationship. The Granger causality results are therefore treated as hypothesis-generating but not as an indication that Western and Asian AI publication trajectories are interdependent.

\bsec{6. Conclusions}%
This study was designed to examine and comparatively analyse the trajectories of global countries dominating AI publications between 2000 and 2025. By using the percentage of global AI publications by country as the core metric, the research quantified a dramatic transition in the global AI research landscape. The core finding is the decisive ascendancy of China, which increased its share of global AI publications from less than 5\% in 2000 to nearly 36\% by 2025, making China the single most dominant contributor to global AI research volume and significantly outpacing any other country or bloc. This study demonstrates a substantial net decline in the relative shares of the historic established players, although the US share recovered partially between 2010 and 2020. Both the US and the EU27 have seen their individual shares reduced by over half, which indicates that their output growth has failed to keep pace with the global expansion of AI research driven largely by Asian economies. In this study, the substantial growth of India\textquoteright{}s contribution alongside China\textquoteright{}s leading position suggests a broader geographical shift of AI publication output towards Asia, in which publication activity is increasingly distributed across several Asian research systems.

Beyond the redistribution of output, the concentration and convergence analyses show that AI publication output among the 12 units has become more concentrated and more dispersed since the mid-2010s, a change driven predominantly by the rise of China, and the $\beta$-convergence regression finds no statistically significant convergence over the full period. The comparison of publication volume and citation impact extends this picture further. China\textquoteright{}s share of high-impact publications exceeds its share of overall publications in 2025, and its average citations per paper has surpassed that of the US. These empirical outputs indicate that China over-performs on citation impact relative to its publication volume, which challenges the general assumption that Western countries retain dominance in highly cited AI scholarship. It should be emphasised, however, that whether this circumstance suggests China\textquoteright{}s dominance in research quality cannot be established from citation-based proxies alone.

If, as argued in the Introduction, academic publication output acts as a leading indicator of technological capability, the transition in publication shares documented here would call for a re-evaluation of AI-related research funding and collaboration policies across Western powers and the EU bloc. This study does not, however, examine the link between publication shares and strategic or geopolitical capability, and its implications for the geopolitical balance should therefore be read as propositions for further research, not as empirical conclusions. The Granger causality outputs, which indicate that the past trajectory of the US share carries limited predictive information for China\textquoteright{}s subsequent trajectory, are similarly treated as hypothesis-generating, and they do not establish interdependence between Western and Asian research systems.

This study has several limitations that should be acknowledged. First, the classification of AI publications relies on field-of-study tagging that follows the OECD standards. Papers in adjacent fields such as natural language processing or computer vision are only included if they are concurrently classified under core AI or machine learning categories, which means the datasets used in this study likely undercount the full breadth of AI-relevant scholarship. Second, the Granger causality analysis identifies only two statistically significant predictive relationships, neither of which remains statistically significant after correction for multiple comparisons, and the analysis does not explain why past values of one country\textquoteright{}s publication share predict another\textquoteright{}s. Third, publication figures for 2025 should be interpreted with the acknowledgement that the datasets used in this study may not fully cover publications from late 2025 due to indexing delays, and the 2025 data should therefore be seen as provisional but not complete. Fourth, France, Germany and Spain are counted both individually and within the EU27 in the inferential analyses, and although the sensitivity check reported in Section 4 indicates that the increase in concentration and dispersion since the mid-2010s is robust to this overlap, the comparison of dispersion between 2000 and 2025 is not. Fifth, the high-impact classification and the average citations per paper are citation-based indicators of influence, which cannot by themselves establish differences in research quality between countries.

These limitations point to three future research directions. Studies employing a broader operational definition of AI would establish how sensitive the trajectories reported here are to taxonomic choice. Qualitative and mixed-methods work on national research policy would help explain the mechanisms underlying the predictive relationships identified by the Granger causality analysis. Finally, co-authorship network analysis would clarify whether the concentration documented in this research paper is accompanied by a comparable concentration of international research collaboration.

\bsec{Data and Code Availability}%
The datasets used in this study are publicly available from the OECD.ai Observatory at \url{https://oecd.ai/en/openalex}, and the Python and STATA code that reproduces Tables 1 to 7 and Figures 1 to 12 is available from J.H. upon reasonable request.

\bsec{Compliance with Ethical Standards}%
J.H. has no potential conflicts of interest to declare. This study only conducts secondary data analysis of open-source panel datasets, so no human participants or animals are involved, and no informed consent needs to be collected to conduct this study.

\bsec{Declaration of generative AI}%
During the preparation of this work J.H. used Claude Code to help with data visualisation of Figures 8-12, alongside organising his thought process when developing this research paper. After writing up this research paper, Claude was used again to check for grammar and coherence.

\bsec{References}%
\begingroup\fontsize{9}{12}\selectfont\setlength{\parskip}{4pt}\everypar{\hangindent=0.25in\hangafter=1}
\noindent
Abbonato, D., Bianchini, S., Gargiulo, F., \& Venturini, T. (2024). Interdisciplinary Research in Artificial Intelligence: Lessons from COVID-19. Quantitative Science Studies, 5(4), 922-35. \url{https://doi.org/10.1162/qss_a_00329}.

Aldasoro, I., Gambacorta, L., Korinek, A., Shreeti, V., \& Stein, M. (2024). Intelligent Financial System: How AI Is Transforming Finance. BIS Working Papers No. 1194. Monetary and Economic Department of the Bank for International Settlements. \url{https://www.bis.org/publ/work1194.pdf}.

Andrews, D. (1993). Tests for Parameter Instability and Structural Change with Unknown Change Point. Econometrica, 61(4), 821\textendash{}56. \url{https://www.jstor.org/stable/2951764}.

Bai, J., \& Perron, P. (1998). Estimating and Testing Linear Models with Multiple Structural Changes. Econometrica, 66(1), 47\textendash{}78. \url{https://doi.org/10.2307/2998540}.

Bai, J., \& Perron, P. (2003). Computation and Analysis of Multiple Structural Change Models. Journal of Applied Econometrics, 18(1), 1\textendash{}22. \url{https://doi.org/10.1002/jae.659}.

Balland, P., Di Girolamo, V., Benoit, F., Ravet, J., \& Hobza, A. (2025). Divided We Fall Behind: Why a Fragmented EU Cannot Compete in Complex Technologies. Publications Office of the European Union. \url{https://doi.org/10.2777/8548441}.

Barro, R., \& Sala-i-Martin, X. (1992). Convergence. Journal of Political Economy, 100(2), 223\textendash{}51. \url{https://doi.org/10.1086/261816}.

Bertelsmann Foundation (2023). Infographic: AI Research and Development in the U.S., EU and China. Retrieved from \url{https://www.bfna.org/digital-world/infographic-ai-research-and-development-in-the-us-eu-and-china-4mk29rb8ig/} (Accessed September 21, 2025).

Bruce, A., Timmerman, L., Fiakpui, N., Lessey, L., Beardah, M., Daeid, N., \& Menard, H. (2025). A Scientometric Review of Explosives Research: Challenges and Opportunities. Forensic Science International, 373: 112513. \url{https://doi.org/10.1016/j.forsciint.2025.112513}.

Carchiolo, V., \& Malgeri, M. (2024). Navigating the AI timeline: From 1995 to today. In Proceedings of the 13th International Conference on Data Science, Technology and Applications (DATA 2024) (pp. 577\textendash{}84). SciTePress. \url{https://doi.org/10.5220/0012856700003756}.

Carlo, A. (2021). Artificial Intelligence in the Defence Sector. In Mazal, J., Fagiolini, A., Vasik, P., \& Turi, M. (Eds.), Modelling and Simulation for Autonomous Systems. Modelling and Simulation for Autonomous Systems (MESAS) 2020. Lecture Notes in Computer Science, 12619 (pp. 269\textendash{}78). Springer, Cham. \url{https://doi.org/10.1007/978-3-030-70740-8_17}.

Choi, C. \& Yoon, J. (2025). AI Policy in Action: The Chinese Experience in Global Perspective. Journal of Policy Studies, 40(2): 1-23. \url{https://doi.org/10.52372/jps.e685}.

Chow, G. (1960). Tests of Equality Between Sets of Coefficients in Two Linear Regressions. Econometrica, 28(3), 591\textendash{}605. \url{https://doi.org/10.2307/1910133}.

Culbert, J., Hobert, A., Jahn, N., \& Mayr, P. (2025). Reference Coverage Analysis of OpenAlex Compared to Web of Science and Scopus. Scientometrics, 130, 2475\textendash{}92. \url{https://doi.org/10.1007/s11192-025-05293-3}.

Ellis, D. (2025). New Report Shows China Dominates in AI Research\textemdash{}and Is Western World's Leading Collaborator on AI. Digital Science. Retrieved from \url{https://www.digital-science.com/blog/2025/07/new-report-shows-china-dominates-in-ai-research/} (Accessed September 20, 2025).

F\"{a}rber, M., \& Tampakis, L. (2024). Analysing the Impact of Companies on AI Research Based on Publications. Scientometrics, 129, 31\textendash{}63. \url{https://doi.org/10.1007/s11192-023-04867-3}.

Gao, X., Wu, Q., Liu, Y., \& Yang, R. (2024). Pasteur's Quadrant in AI: Do Patent-Cited Papers Have Higher Scientific Impact? Scientometrics, 129(2), 909\textendash{}32. \url{https://doi.org/10.1007/s11192-023-04925-w}.

Gerlich, M. (2024). Brace for Impact: Facing the AI Revolution and Geopolitical Shifts in a Future Societal Scenario for 2025\textendash{}2040. Societies, 14(9): 180. \url{https://doi.org/10.3390/soc14090180}.

Gohil, M. (2023). A Study on the Impact of Artificial Intelligence (AI), Automation and the Digital Transformation on Society. Revista Review Index Journal of Multidisciplinary, 3(4), 22-7. \url{https://doi.org/10.31305/rrijm2023.v03.n04.004}.

Granger, C. (1969). Investigating Causal Relations by Econometric Models and Cross-Spectral Methods. Econometrica, 37(3), 424\textendash{}38. \url{https://doi.org/10.2307/1912791}.

Hamilton-Hart, N. \& Yeung, H. (2021). Institutional Under Pressure: East Asian States, Global Markets and National Firms. Review of International Political Economy, 28(1), 11-35. \url{https://doi.org/10.1080/09692290.2019.1702571}.

Hong, S., Zhong, D., \& Um, K. (2025). The Impact of Artificial Intelligence (AI) Adoption on Operational Performance in Manufacturing. Journal of Manufacturing Technology Management, advance online publication, 1-23. \url{https://doi.org/10.1108/JMTM-03-2025-0227}.

Hood, W. W., \& Wilson, C. S. (2001). The Literature of Bibliometrics, Scientometrics, and Informetrics. Scientometrics, 52, 291\textendash{}314. \url{https://doi.org/10.1023/A:1017919924342}.

Jurowetzki, R., Hain, D., Mateos-Garcia, J., \& Stathoulopoulos, K. (2021). The Privatization of AI Research(-ers): Causes and Potential Consequences\textemdash{}From University-Industry Interaction to Public Research Brain-Drain? arXiv:2102.01648. \url{https://doi.org/10.48550/arXiv.2102.01648}.

Jurowetzki, R., Hain, D., Wirtz, K., \& Bianchini, S. (2025). The Private Sector is Hoarding AI Researchers: What Implications for Science? AI \& Society, 40, 4145\textendash{}52. \url{https://doi.org/10.1007/s00146-024-02171-z}.

Kendall, M. (1975). Rank Correlation Methods (4th ed.). Charles Griffin.

Khan, M., Khan, H., Omer, M., Ullah, I., \& Yasir, M. (2025). Impact of Artificial Intelligence on the Global Economy and Technology Advancements. In El Hajjami, S., Kaushik, K., \& Khan, I. (Eds.), Artificial General Intelligence (AGI) Security. Advanced Technologies and Societal Change (pp. 147\textendash{}80). Springer, Singapore. \url{https://doi.org/10.1007/978-981-97-3222-7_7}.

LeCun, Y., Bengio, Y., \& Hinton, G. (2015). Deep Learning. Nature, 521, 436\textendash{}44. \url{https://doi.org/10.1038/nature14539}.

Liu, N., Shapira, P., \& Yue, X. (2021). Tracking Developments in Artificial Intelligence Research: Constructing and Applying a New Search Strategy. Scientometrics, 126(4), 3153\textendash{}92. \url{https://doi.org/10.1007/s11192-021-03868-4}.

Mahdi, S., Battineni, G., Khawaja, M., Allana, R., Siddiqui, M., \& Agha, D. (2023). How Does Artificial Intelligence Impact Digital Healthcare Initiatives? A Review of AI Applications in Dental Healthcare. International Journal of Information Management Data Insights, 3(1): 100144. \url{https://doi.org/10.1016/j.jjimei.2022.100144}.

Mann, H. (1945). Nonparametric Tests Against Trend. Econometrica, 13(3), 245\textendash{}59. \url{https://doi.org/10.2307/1907187}.

Merton, R. (1968). The Matthew Effect in Science. Science, 159(3810), 56\textendash{}63. \url{https://doi.org/10.1126/science.159.3810.56}.

Mingers, J. \& Leydesdorff, L. (2015). A Review of Theory and Practice in Scientometrics. European Journal of Operational Research, 246(1), 1-19. \url{https://doi.org/10.1016/j.ejor.2015.04.002}.

OECD.ai Observatory (n.d.). OpenAlex: Methodological Note. Retrieved from \url{https://oecd.ai/en/openalex} (Accessed November 21, 2026).

Perianes-Rodr\'{\i}guez, A., Waltman, L., \& van Eck, N. (2016). Constructing Bibliometric Networks: A Comparison Between Full and Fractional Counting. Journal of Informetrics, 10(4), 1178\textendash{}95. \url{https://doi.org/10.1016/j.joi.2016.10.006}.

Podda, L. (2025). China's Drive to Dominate the AI Race. Atlas Institute for International Affairs. Retrieved from \url{https://atlasinstitute.org/chinas-drive-to-dominate-the-ai-race/} (Accessed September 20, 2025).

Priem, J., Piwowar, H., \& Orr, R. (2022). OpenAlex: A Fully-Open Index of Scholarly Works, Authors, Venues, Institutions, and Concepts. arXiv:2205.01833. \url{https://doi.org/10.48550/arXiv.2205.01833}.

Quimba, F. \& Barral, M. (2024). ASEAN Centrality amid Increasing Global Multipolarity. PIDS Discussion Paper Series, No. 2024-38. Philippine Institute for Development Studies. \url{https://doi.org/10.62986/dp2024.38} (Accessed September 20, 2025).

Rhoades, S. (1993). The Herfindahl-Hirschman Index. Federal Reserve Bulletin, 79(3), 188\textendash{}9. Retrieved from \url{https://fedinprint.org/item/fedgrb/15160} (Accessed September 17, 2026).

Sala-i-Martin, X. (1996). The Classical Approach to Convergence Analysis. The Economic Journal, 106(437), 1019\textendash{}36. \url{https://doi.org/10.2307/2235375}.

Sen, P. (1968). Estimates of the Regression Coefficient Based on Kendall's Tau. Journal of the American Statistical Association, 63(324), 1379\textendash{}89. \url{https://doi.org/10.1080/01621459.1968.10480934}.

Sinha, A., Shen, Z., Song, Y., Ma, H., Eide, D., Hsu, B., \& Wang, K. (2015). An Overview of Microsoft Academic Service (MAS) and Applications. In Proceedings of the 24th International Conference on World Wide Web (WWW \textquoteright{}15 Companion) (pp. 243\textendash{}6). ACM. \url{https://doi.org/10.1145/2740908.2742839}.

Stanford University Human-Centered Artificial Intelligence (HAI) (n.d.). The Global AI Vibrancy Tool. Retrieved from \url{https://hai.stanford.edu/ai-index/global-vibrancy-tool} (Accessed September 20, 2025).

Wang, K., Shen, Z., Huang, C., Wu, C., Eide, D., Dong, Y., Qian, J., Kanakia, A., Chen, A., \& Rogahn, R. (2019). A Review of Microsoft Academic Services for Science of Science Studies. Frontiers in Big Data, 2: 45. \url{https://doi.org/10.3389/fdata.2019.00045}.

\endgroup

\end{document}